
\input harvmac
\Title{\vbox{\baselineskip12pt\hbox{IC/94/390}}}
{\vbox{\centerline{Non-Abelian Bosonization and Higher Spin Symmetries}}}

\centerline{Raiko P. Zaikov\footnote{*}{Permanent
address since December 16, 1994; Institute for Nuclear Research
and Nuclear Energy, Boul. Tzarigradsko Chaussee 72, 1784 Sofia,
e-mail: zaikov@bgearn.bitnet} \footnote{$^\star$} {Supported by Bulgarian
Foundation on Fundamental Research under contract Ph-318/93-96}}
\bigskip\centerline{International Centre for Theoretical Physics}
\centerline{P.O. Box 586, 34100 Trieste, Italy}

\vskip .3in

\noindent{\bf Abstract. } The higher spin properties of the
non-abelian bosonization in the classical theory are investigated.
Both the symmetry transformation algebra and the classical
current algebra for the non-abelian free fermionic model are
linear Gel'fand-Dickey type algebras.  However, for the
corresponding WZNW model these algebras are different. There
exist symmetry transformations which algebra remains the linear
Gel'fand-Dickey algebra while in the corresponding current
algebra nonlinear terms arised. Moreover, this algebra  is
closed (in Casimir form) only in an extended current space in
which nonlinear currents are included. In the affine sector, it
is necessary to be included higher isotopic spin current too. As
result we have a triple extended algebra.

\Date{11/94}


\newsec{Introduction}

\noindent The non-abelian free fermionic model and the WZNW model
are not distinguished usually from some authors because as was
shown by Witten
\ref\rW{E.\ Witten, Comm. Math. Phys., 92 (1984) 455.} the
$SO(N)$ non-abelian free fermionic model and the corresponding
WZNW model are equivalent on a classical as well as a quantum
level (see also \ref\Pol{A.\ A.\ Polyakov and P.\ B.\ Wiegmann,
Phys. Lett. {\bf 131B} (1993) 121: {\bf 141B} (1984) 223.}).
This equivalence is understood as the presence of the same ($SO(N)\otimes
SO(N)$,
conformal and affine Kac-Moody) symmetry. This symmetry involves
coinciding current algebras on a classical level as well on a
quantum level and moreover one have identical anomalies for both
models in the quantum case.  Recently the Witten bosonization
procedure was checked more directly deriving the WZNW model from
the gauged free fermionic model \ref\rDNS{P.H. Damgaard, H.B.
Nielsen and R. Sollacher, Nucl. Phys., {\bf B 385} (1992) 227;
{\bf B 414} (1994) 541; Phys. Let., {\bf 296 B} (1992) 132; {\bf
322 B} (1994) 131.},
\ref\rBQ{C.\ P.\ Burgess and F.\ Quevedo, Nucl. Phys. {\bf B 421}
(1994) 373; Phys. Lett., {\bf 329 B} (1994) 457}
 in the path-integral approach.

The currents for both the free fermionic model and  WZNW model
are chiral, i.e.
\eqn\eii{\partial _{\bar z}J=\partial _z\bar J=0,}
which shows that, besides the conformal and affine Kac-Moody
symmetries, these models admit higher spin extensions
known as $W$--symmetry.  Usually, the Drinfeld-Sokolov reduction
procedure  is applied
\ref\rDS{Drinfeld, V. and Sokolov, V., J. Sov.  Math., 30 (1984)
1975.}\nref\rFL{Fateev, V. A. and Lukyanov, S. L., Int. J. Mod. Phys.,
A3 (1988) 507 \semi Sov. Phys. JETP, 67 (1988) 447 \semi Sov.
Sci. Rev. A Phys., {\bf 15}(1990) 1.}
\nref\rBFRFW{J. Balog, L. Feh\' er, L. O'Raifertaigh, P. Forgas
and A. Wipf, Ann. Phys. (N. Y.), {\bf 203} (1990)
76.} \nref\rY{K. Yamaguchi, Phys.  Lett., {\bf B259} (1991) 436.}
\nref\rYu{F. Yu and Y. -S. Wu, Phys. Let. {bf B263} (1991)
220.}--\ref\rBC{L. Bonora and C.S.  Xiong, Phys. Lett., {\bf
B285} (1992) 191 \semi R. Paunov, Phys. Lett., {\bf B309}
(1993) 297.} to
investigate of the $W$--symmetry in the models with chiral
currents (for complete list of references see \ref\rBS{P.
Bouknegt and K. Schoutens, Phys.  Reports, {\bf 223} (1993)
183.} and \ref\rFRRTW{F. Feh\' er, L. O'Raifertaigh, P. Ruelle,
I. Tsutsui and A. Wipf, Phys. Reports, {\bf 222} 91992) 1.}).
Despite the fact that the higher spin currents obtained by this method
are independent, they are connected to the choice of the
subgroup for reduction.  Hence, these currents do not
possess for completeness in general.  Moreover, some of the characteristic
properties of the model under consideration can be lost,
because we start from the affine Kac-Moody algebra and the
fields transformation laws usually are not analized.\foot{One
attempt to relate the Drinfeld--Sokolov procedure to the
Lagrangian approach is discussed in the preprint \ref\rDSTS{A.
Deckmyn, R. Siebelink, W. Troost and A. Sevrin, {\it On the
Lagrangian realization of non-critical {\cal W}-strings},
Preprint KUL-TF-94/38 and CERN-TH.7477/94; hep-th/9411221.}
appearing after completion of the present paper.} The properties
of the examined model become more transparent when the hidden
symmetry method is applied \rBS . This method consists in
searching for additional symmetries of the action which generate
higher spin conserved quantities \ref\rD{L. Dolan and A. Roos,
Phys. Rev.  {\bf D22} (1980) 2018.}, \ref\rZai{R. P.  Zaikov,
Theor. Math.  Phys., {\bf 53} (1982) 534.}. By means of this
method in the papers
\ref\rZa{R. P.  Zaikov, {\it Extended non-abelian Symmetries for
Free Fermionic Model}, preprint ICTP, IC/93/237, Trieste 1993,
hep-th/9308016.} and \ref\rZ{R. P. Zaikov, Lett. Math. Phys.,
{\bf 32} (1994) 283.}
some difference between the symmetry properties of
the classical free fermionic model and the classical WZNW model
with respect to the higher spin transformations was obtained.  Indeed,
although  the action integrals of both models are invariant
with respect to the transformations satisfying isomorphic
extended symmetry algebras; the \ $DOP(S^1)$ \ transformations
\ref\rRad{A. O.  Radul, Pis'ma Zh. Ehsp. Teor. Fis. , {\bf 50}
(1989) 341\semi Phys. Lett. {\bf B265} (1991) 143.}
\nref\rPop{C. N. Pope, L. J. Romans and X. Sheng, Phys. Lett.
{\bf B336} (1990) \semi Nucl. Phys., {\bf B265} (1991) 86.}
--\ref\rB{E. Bergshoeff, C. N. Pope, L. J. Romans, E. Sezgin and
X. Shen, Phys. Lett., {\bf B245} (1990) 447.}, the corresponding
higher spin currents for both models have different properties
\ref\rAA{Abdalla, E., Abdalla, M., C., B., Sotkov, G. and
Stanishkov, \ {\it "Of critical current algebra"} \ preprint
IFUSP/1027/93, Sao Paulo 1993; hep-th/9302002.}, \rZa \ and \rZ .
In the paper \rPop , \rYu , \rAA \ it is shown that the higher
spin fermionic currents satisfied a closed current algebra,
which coincides with the algebra of the corresponding field
transformations \rZa .  Really, applying the Witten non-abelian
bosonization procedure \rW
\  the corresponding bosonic currents and the
 field transformations generating these currents were obtained in \rZ .
Moreover, it was shown that these infinitesimal bosonic field
transformations being the symmetry of the WZNW  action satisfied
the same algebra as the corresponding fermionic field
transformations. However, the transformation laws for the higher
spin bosonic currents essentially differ from the corresponding
fermionic currents law. This difference consists in that the
higher spin fermionic currents form invariant space, while the
corresponding bosonic currents do not span an invariant space .
To obtain an invariant bosonic currents space the higher spin
Noether currents were complemented by additional nonlinear
higher spin currents
\ref\rRZ{R. P. Zaikov, {\it "Hidden symmetry approach to the
extended symmetries in non-abelian free fermionic and WZNW
models"}, Proc. of XXVII$^{th}$ Symposium on the Theory of
Elementary Particles, Wendisch-Rietz, September 7-11, 1993.}.

The present paper is devoted to the more systematic
investigation of the properties of the Witten non-abelian
bosonization procedure with respect to the higher spin symmetry
within the hidden symmetry approach. In the second section a
brief review of the results of the papers \rZa \  is given. The
explicit form of the infinitesimal fermionic fields
transformation Lie algebra and the corresponding classical
current transformation laws, i.e. current algebra  are presented
in the form convenient for comparison with those obtained for
the WZNW model. These higher spin $SO(N)$ algebras are closed
only if they are completed to $GL(N)$ algebras.  It is shown
also that the correspondence between the symmetry
transformations and the currents is one to one up to the terms
containing derivatives from lower spin quantities.

In the third section the corresponding higher spin bosonized
currents and the corresponding  field transformations are
considered. The infinitesimal transformations being  on-shell
symmetry of the classical WZNW action satisfied the same linear
Gel'fand-Dickey algebra \ref\rGD{Gel'fand, I. M and Dickey, L.
A., {\it "A Family of Hamiltonian Structures Connected with
Integrable Non-Linear Differential Equations"}, IPM preprint, AN
SSSR, Moskow 1978.} as the fermionic field transformations.
However, in this case the correspondence between the field
transformations and the conserved currents is not one to one. For
instance, from the bilinear higher spin currents which
correspond to the linear \ $DOP(S^1)$ \ field transformations
applying the Poisson brackets we derive the transformations which
differ from the initial transformations by nonlinear terms.
These nonlinear transformations generate nonlinear higher spin
currents involving transformations containing higher
nonlinearity and so one.  For instance, for any  current with
spin \ $s=m$ \ we can obtain $m-1$ new nonlinear currents with
the same spin.

In the fourth section
the classical current algebra for $O(N)$ WZNW
model is obtained. The higher spin $U(1)$ current algebra
($W_\infty $ algebra) is closed  (in terms of Casimir currents)
only if to the spin \ $s=m$ \ bilinear current a new \ $m-1$ \
additional nonlinear currents with the same spin are included.
In the case of $SO(N)$ isotopic currents with higher conformal
spin we have more complicated situation. In this case to have
closed current algebra the initial currents must be completed
with $GL(N)$ isotopic tensor currents too. As a result a triple
extended affine current algebra is obtained. For the abelian
case, which also is considered, the obtained algebra can be
considered as nonlinear $\widehat W_{1+\infty }$ algebra
too \ref\rYW{F. Yu and Y. -S. Wu, J. Math. Phys. {\bf 34} (1993)
5851 \semi Phys. Let., {\bf B263} (1991) 220; {\bf B294} (1992)
177 \semi Nucl. Phys., {\bf B373} (1992) 713 \semi Phys. Rev.
Lett., {\bf 68} (1992) 2996.}. However, for the non-abelian case
in general we have nonlinear closed algebra (without adding
new terms) only in terms of matrix valued currents \rAA ,
\ref\rBil{A.  Bilal, {\it Multi-component KdV hierarchi,
V-algebra and non-abelian Toda theory}, hep-th/9401167.}.

In the fifth section some concluding remarks are given.

In the Appendix the explicit form of the unconstrained higher
spin currents is given for the $SL(2,R)$ WZNW model in terms of
the Gauss decomposition components field.


\newsec{Higher Spin Field Transformations for Free Fermionic Model}

\noindent As is well known, always when we are dealing with a
Lagrangian theory whose action is invariant with respect to some
$N$-parametric rigid (gauge) group transformations $G$ with Lie
algebra ${\cal G}$ we know from the Noether theorem that (for
the application of the Noether theorem to the conformal theories
see
\ref\rRaf{L.  O'Raifertaigh, {\it Short recall of
two-dimensional conformal theories}, Talk given at IVth Regional
Conference on Mathematical Physics, Tehran 1990, DIAS preprint,
STP-90-42.}):

{\it i})  We have $N$ conserved quantities,

{\it ii})  These quantities act as generators of the group $G$,

{\it iii})  The corresponding current algebra coincides with the Lie
algebra ${\cal G}$ of the group $G$.

\noindent It is evident that the assertion {\it iii}) is a
consequence of {\it ii}). These assertions take place also for
some infinite-parametric group such as conformal group in
two-dimensions, affine Kac-Moody transformations and for the
higher spin extended transformations as the area-preserving
transformations  $w_\infty $, extended conformal and affine
Kac-Moody transformations in the case of free fermionic model, e.t..
However, as was shown in \rZ \ (see also \rRZ ) a class
transformations exist for the case of the WZNW model in which
terms violating the assertions
\  {\it ii}) and {\it iii}) appear. The latter means that
the current$\Longleftrightarrow $symmetry correspondence is not
invertible in this case. Consequently, applying the Poisson
bracket we do not return to the same symmetry transformation and
the corresponding current algebra does not coincide with the
Lie algebra of the initial infinitesimal fields transformations.
In the present paper we consider two examples for comparison:
the non-abelian free fermionic model for which \ {\it i) - iii)}
\ are satisfied for higher spin extended symmetry
transformations and the WZNW model for which \ {\it ii)} \ and \
{\it iii)}
\ are violated on the higher spin level.

In the present paper we consider transformations which are
on-shell symmetry too, i.e.  transformations leaving the action unchanged
only on the solutions of the equations of motions. We
recall that such kinds of symmetries usually appear in the
supersymmetric theories. According to the modified
Noether theorem given by Ibraghimov
\ref\rI{N.\ H.\ Ibraghimov, Theor. Math. Phys.,{\bf 1} (1969) 350.},
\ref\rCPV{E.\ Candotti, C.\ Palmer and B.\ Vitale, Nuovo Cim., {\bf 70A}
(1970) 233.}  to any continuous one parametric on-shell symmetry
transformation there also corresponds one conserved quantity, i.e. the
assertion {\it i}). As we will see below the assertions {\it ii) {\rm and}
iii)} can be violated on the higher spin symmetry level.


 First we reconsider the classical free fermionic model whose
higher spin conserved currents are given by:
\eqna\ea
$$
\eqalignno{{\cal V}^{n}&=\bar \psi \partial _z^{n+1}\psi ,&\ea a\cr
{\cal J}^n_a&=\bar \psi t_a\partial _z^{n}\psi , \qquad \
(n=0,1,\dots &\ea b\cr }
$$
being a higher spin extension of the holomorphic
components of the stress-energy tensor, \ ${\cal V}^0=T_{zz}$
and of the $SO(N)$ or $SU(N) (GL(N))$ (for definiteness) isotopic
current \ ${\cal J}^0_a=J_a$ respectively. In the formulas
\ea{} \ $z$ denotes a light-cone variable $x_+=(x_0+x_1)/2$ or the
corresponding complex variable, $\psi $ is a holomorphic
component of the free spinor field and \ $t_a\in {\cal G}$ \ are
the generators of the group \ $G$ \ in fundamental
representation. We restrict our considerations only to the
holomorphic components $\partial _{\bar z}{\cal V}=0$ , having
in mind that the anti-holomorphic components have a similar form
and we are concerned with the $SO(N)$  model for which $\psi $ is a
holomorphic component of a anticommuting Majorana spinor , i.e.
$\bar \psi =\psi $. Notice, that in \ea{} and what follows the
currents normalization constants are omitted.

When we are dealing with the complex spinor field all the
currents \ea{}  are independent, however, for the Majorana
spinor field this is not the case. Indeed, using the
formula \rZa :
\eqn\el{\partial ^m\phi \Omega \chi =\sum _{p\ge 0}{m\choose p}
\partial ^p(\phi \Omega \partial ^{m-p}\chi ),}
where $\Omega $ is a constant matrix one can get:
\eqn\ela{\eqalign{{\cal V}^1&=\partial {\cal V}^0, \cr
{\cal V}^3&=2\partial {\cal V}^2-\partial ^3{\cal V}^0, \cr
{\cal V}^5&=3\partial {\cal V}^4-5\partial ^3{\cal V}^2+
3\partial ^5{\cal V}^0, \cr
&\vdots \cr }}
for symmetric matrix $\Omega $ and
\eqn\ele{\eqalign{
{\cal J}_a^1&={1\over 2}\partial {\cal J}_a^0, \cr
{\cal J}_a^3&={3\over 2}\partial {\cal J}_a^2-
{1\over 4}\partial ^3{\cal J}_a^0, \cr
{\cal J}_a^5&={5\over 2}(\partial {\cal J}_a^4
-\partial ^3{\cal J}_a^2)+{1\over 2}\partial ^5{\cal J}_a^0, \cr
& \vdots \cr  }}
for skew symmetric matrix $\Omega $.
For arbitrary odd spin $U(1)$ current \ea{a} and even spin
isotopic vector current \ea{b} we obtain respectively:
\eqna\elb
$$
\eqalignno{{\cal V}^{2m+1}&=\sum _{p=0}^mC^m_p\partial
^{2p+1}{\cal V}^{2(m-p)},&\elb a\cr
{\cal J}_a^{2m+1}&=\sum _{p=0}^m\widetilde C^m_p\partial
^{2p+1}{\cal J}_a^{2(m-p)},&\elb b\cr }
$$
where $C^m_p$ are constants which could be derived from \el .
The formulas \ela \ and \elb{} show us that by a suitable choice
of the constants $R^n_p$ in formula
\eqn\eazz{W^n\rightarrow \widehat W^n=\sum _{p=0}^nR^n_p\partial
^pW^{n-p}}
we can obtain a basis in which \ ${\cal V}^{2m+1}\equiv 0 \ {\rm
and} \ {\cal J}_a^{2m+1}\equiv 0$. For
convenience we use the basis \ea{} in which the depending
currents are not excluded, hence the relations
\elb{}  must be kept in mind.

The currents \ea{}  are Noetherian, i.e.\ they follow from the
Noether theorem as a consequence of the on-shell symmetry of the
free fermionic action with respect to the infinitesimal
transformations:
\eqna\eas
$$
\eqalignno{\delta ^{m}\psi(z)&=k_m(z)\partial
^{m+1}\psi ,&\eas a\cr
\widehat\delta ^{m}\psi(z)&=\alpha ^a_m(z)t_a\partial
^{m}\psi ,&\eas b \cr }
$$
where \ $k_m(z)$ \ and \ $\alpha ^a_m(z)$ are arbitrary
holomorphic functions. It is easy to check that the
transformations \eas{}   satisfied the
following Lie algebra:
\eqna\etr
$$
\eqalignno{[\delta ^m(k),
\delta ^n(h)]\psi (z)=
\sum _{r\ge 0}& \delta ^{m+n-r+1}
\Bigl([h_{n+1},k_{m+1}]^r_-\Bigr)\psi (z),&\etr a \cr
[\delta ^m(k),\widehat \delta ^n(\hat \beta )]\psi (z)
=\sum _{r\ge 0} & \widehat \delta ^{m+n-r+1}([\hat \beta _n
,k_{m+1}]^r_-)\psi (z),&\etr b \cr
[\widehat \delta ^m(\hat \alpha ),
\widehat \delta ^n(\hat \beta )]\psi (z)=
\sum _{r\ge 0} &\biggl(\widehat \delta
^{m+n-r}_c(f_{ab}^c[\beta ^b_n,\alpha ^a_m]^r_+t_c) \cr
&  +\widehat \delta
^{m+n-r}([\beta ^b_n,\alpha ^a_m
]^r_-t_{ab})
\biggr)\psi (z),&\etr c \cr }
$$
where $t_{ab}=\{t_a,t_b\}/2$,
\eqn\efda{2\lbrack \beta _n,\alpha _m\rbrack ^r_{\pm }=
\pmatrix{ n \cr r \cr }\beta _n\partial ^r\alpha _n\pm
\pmatrix{ m \cr r \cr }\alpha _m\partial ^r\beta _n}
and it is taken into account that the binomial coefficients
\ $\pmatrix{m \cr r \cr }=0$ \ for \  $r>m$.
This algebra contains as subalgebras the Virasoro algebra, the
$W_{\infty }$ algebra, semi-direct product of the Virasoro and
the affine Kac-Moody algebras.

{\bf Observation:} If $m\ne n$
in the r.h.s. of \etr{c} \  the anticommutators of the matrix
generators of $SO(N)$ are also included. The origin of these
anticommutators is the following: When $m\ne n$ in the
commutator of two transformations \eas{b} \ the matrix products
$t_at_b$ and $t_bt_a$ appears by different coefficients, so
that they do not form the Lie product only. Indeed, taking into account
that the product \ $t_at_b$ \ can always be represented as
$$
t_at_b={1\over 2}[t_a,t_b]+{1\over 2}\{t_a,t_b\}
$$
and that $\{t_a,t_b\}\in {\cal U}(N)\bigl({\cal GL}(N)\bigr)$ when
$t_a\in {\cal SU}(N)\bigl({\cal SL}(N)\bigr)$ then $t_at_b\in {\cal
U}(N)\bigl({\cal GL}(N)\bigr)$. The latter allows us to
conclude that the extended affine $SU(N)$ Kac-Moody algebra
is not closed alone, because the affine $U(1)$
transformations coincides with the $W$--transformations. However, if
$t_a\in {\cal SO(N)}$ then the anticommutator $\{t_a,t_b\}\notin
{\cal SO(N)}$ for any nontrivial representation of
$SO(N)$. The latter is a consequence of the uncompleteness of
the space of matrices \ $t_a$ \ in any
representation of $SO(N)$. In this case $t_at_b\in {\cal GL(N)}$.
Indeed the $N(N-1)/2$ \ skew symmetric $N\times N$ \ matrix
together with $N(N+1)/2$ symmetric matrices $\{t_a,t_b\}$ \ form a
complete $N\times N$ \ real matrix space.

According to the second part of the Noether theorem (assertion
{\it ii}) applying the Poisson bracket we derive the following
field transformations laws: \eqna\eaa
$$
\eqalignno{\delta ^{l+1}\psi (z) &
=\int dxk_l(x)\{{\cal V}^{l+1}(x),\psi(z)\}_{PB}=
\cr &
= -2\sum _{s\ge 0}\delta _{l,2s}k_l\partial ^{l+1}\psi
+ (-)^{l+1}\sum _{r\ge 1}{l+1\choose r}
\partial ^rk_l \partial ^{l-r+1}\psi (z),&\eaa a\cr
\widehat\delta ^{l}\psi (z) &
= \int dx\alpha _l^a(x)\{{\cal J}^{l}_a(x),\psi (z)\}_{PB}=
\cr &
=-2\sum _{s\ge 0}\delta _{l+1,2s}\alpha_l^a\partial ^{l}t_a\psi +
(-)^l\sum _{r\ge 1}{l\choose r}\partial ^r\alpha _l^a
\partial ^{l-r}t_a\psi (z), &\eaa b\cr
\widehat \delta ^{l+1}\psi (z) &
= \int dx\alpha _l^{ab}(x)\{{\cal J}^{l+1}_{ab}(x),\psi (z)\}_{PB}\cr &
=-2\sum _{s\ge 0}\delta _{l,2s}\alpha_ l^{ab}\partial ^{l+1}t_{ab}\psi \cr
&+(-)^{l+1}\sum _{r\ge 1}{l+1\choose r}\partial ^r\alpha
_l^{ab} \partial ^{l-r+1}t_{ab}\psi (z), &\eaa c\cr }
 $$
where
\eqn\euu{{\cal J}^l_{ab}={1\over 2}\psi \{t_a,t_b\}\partial ^{l}\psi .}
Comparing the transformations \eaa{} \ with  \eas{} \ we can
conclude that they differ by the terms \ $\partial ^{m-r}(\delta
^r\psi ), \ (r=0, \dots m-1)$ \ and moreover that the terms \
$h_{l+1}\partial ^{l+1}\psi $ \ are absent for odd $l$ in
\eaa{a,c} and for even \ $l$ in \eaa{b}. In both cases the
corresponding to these transformations currents are not
independent which agree with \elb{}.   This means that applying
both parts of the Noether theorem we return to the same (up to
redefinition) transformation. The absence of some terms  in
formulas
\eaa{} included  in  \eas{}  is a property of the basis under
considerations.  The independent Noether currents corresponding to the
transformations \eaa{}  also coincide up to redefinition to the currents
\ea{}. Another important difference, however, is that the transformations
\eaa{}  are off shell-symmetry of the action, i.e. they leave
the free fermionic action up to surface terms for arbitrary
fields configurations.

It is straightforward to check that the transformations \eaa{} \
 satisfy the Lie algebra which differs from \etr{}  only by the factor
$(-)^r$.

Using \eas{a} we derive the currents transformation law
\eqn\ee{\eqalign{\delta ^m{\cal V}^n & =\sum_{p\ge 1}
{n+1\choose p}\partial ^pk_m{\cal V}^{m+n-p+1} \cr & -
k_m\sum _{q\ge 0}(-)^{m-q}{m+1\choose q}
\partial ^q{\cal V}^{m+n-q+1}, \cr }}
which shows that they span an invariant
space with respect to the transformations \eas{a}. Moreover, it
is easy to check that all the currents \ea{} form invariant
space with respect to the both laws \eas{} and \eaa{}, hence,
they satisfied a closed current algebra.

When $m=0$ the law \ee \ becomes:
\eqn\ess{\delta ^0{\cal V}^n=(n+1)\partial k_0{\cal V}^n+
k_0\partial {\cal V}^n+\sum _{p\ge 2}{n+1\choose p}\partial ^pk_0
{\cal V}^{n-p+1}}
hence $V^n$ is a quasi-primary field.

\newsec{Higher Spin Symmetry in WZNW Model}

\noindent First we consider the abelian case in order to gain
intuition and to make the choice of higher spin bosonic currents
more transparent.

\subsec{Abelian Bosonization}

\noindent According to the abelian bosonization procedure we
should pair a bosonic currents to any fermionic current  \ea{a}
\eqn\eea {V^n(z)={\cal U}^{-1}\partial ^{n+2}{\cal U}(z)=
e^{-\varphi (z)}\partial ^ne^{\varphi (z)},}
where $\varphi $ is dimensionless scalar field. We recall that
from \eea \ it follows:
\eqn\eeb{V^n(z)={\cal P}^{n+2}(J)=\biggl(J(z)+\partial
\biggr)^{n+2}.1,}
where {\cal P}(J) are the Fa\'a di Bruno polynomials \ref\rFB{F.
Fa\'a di Bruno, Quart. Jour. Pure Appl. Math.., {bf 1} (1857)
359}, \ref\rAFGZ{H. Aratyn, L. A. Ferreira, J. F. Gomes and A.
H. Zimerman, Nucl. Phys., {\bf B402} (1993) 85.} and
$J={\cal U}^{-1}\partial {\cal U}=\partial \varphi $.

It is easy to show that the currents \eea \ are generated by the
transformations
\eqn\eeaa{\delta ^m{\cal U}(z)=k_m(x)\partial ^{m+1}{\cal U}(z),}
being an on-shell symmetry of the action
\eqn\eeab{S={1\over 2}\int d^2z\partial _z{\cal U}^{-1}\partial
_{\bar z}{\cal U}.}
Here we consider the $U(1)$ group element ${\cal U}$ as
independent field variables, i.e. we deal with the $U(1)$ chiral
model. Notice, that the
transformations \eeaa \ coincide with corresponding
fermionic ones \eas{a} by the form, hence they  satisfy the same \
$W_{1+\infty }$ \ (linear Gel'fand-Dickey) algebra \etr{a}.

The canonical currents corresponding to the
transformations \eeaa \ are given by
\eqn\eua{V^n_{can}=\partial {\cal U}^{-1}\partial ^{n+1}{\cal U}.}
However, taking into account the identity
\eqn\eub{V^{n+1}=(J+\partial )V^n}
which follows from \eeb , we obtain
\eqn\euc{V^n=\partial V^{n-1}-V^n_{can}.}
Consequently, the currents \eea \  can be considered as
improved ones. Indeed, seting $n=0$ in \euc \ we get the
improved stress-energy tensor
$$
T=-V^0_{can}+\partial j={1\over 2}j^2+\partial j.
$$

According to the second part of the Noether theorem we have:
\eqn\eec{\eqalign{\delta ^n{\cal U}
&
=\int dyk_n(y)\{V^n(y),{\cal U}(z)\}_{PB}
\sum _{p\ge 1}(-)^{p-1}{n\choose p}\partial ^{p-1}k_n
\partial ^{n-p}{\cal U} \cr
& ={\cal U}(z)\sum _{p,r\ge 1}(-)^{p-1}{\cal C}^{n+2,p-1}_{p,r}
\partial ^{p-r-1}k_n(z)\partial ^r\biggl({\cal U}^{-1}
\partial^{n-p+2}{\cal U}(z)\biggr), \cr }}
where
\eqn\ezzr{{\cal C}^{m,n}_{p,q}={m\choose p}{n\choose q}}
and it is taken into account that ${\cal  C}^{m,n}_{p,q}=0$ if
$p>m$ or $q>n$.

{\bf Observation:} In the transformation law \eec \ we see also
a nonlinear part (with respect to ${\cal U}$ in contrast to
the corresponding fermionic case
\eaa {}, and the bosonic transformations \eeaa \ where nonlinear
terms are absent. The presence of these nonlinear terms in \eec
\ is a consequence of the nonlinearity of the bosonization
procedure \eea
\ and they appear when higher order derivatives are included. Another
consequence of this nonlinearity is the noninvariance of the
current space \ \eea \ with respect to the linear
transformations  \eeaa
\ (see \rZ , \rRZ , \rAA  ).

Indeed, applying the linear transformations \eeaa \ we obtain:
\eqn\eecc{\delta ^mV^n=-k_mV^{m-1}V^n+
\sum _{p\ge 0}{n+2\choose p}\partial ^pk_mV^{m+n-p+1}.}
If \ $m,n\ne 0$ \ the first term in the r.h.s. is essentially
nonlinear with respect to the currents ${\cal V}^n$  and it is
not present in the fermionic current transformation laws \ee .
For \ $m=0$ \  the formula \eecc \ becomes
\eqn\eecd{\delta ^0V^n=(n+2)\partial k_0V^n-k_0\partial V^n+
\sum _{p\ge 2}{n+2\choose p}\partial ^pk_mV^{m+n-p+1},}
i.e. the higher spin bosonic currents  \eea \ are quasi-primary
fields too.  However, with respect to the nonlinear
transformations \eec \ only the stress-energy tensor $V^0$ is a
primary field. All this is an indication that we have
nonlinear current algebra.

The currents which can be obtained from the nonlinear transformations \eec \
contain nonlinear terms \
$V^{m-l}V^l, (l=0, \dots m)$. Consequently, by multiple application of
the Noether theorem we obtain the terms with an increasing nonlinearity.
It is easy to check that the product $V^mV^n$ is impossible (if
$m,n>-1$) to be represented linearly in terms of the currents $V^k$
and their derivatives, i.e.
\eqn\egh{V^mV^n\ne \sum _{k=0}^{m+n}A^{m,n}_k\partial ^kV^{m+n-k},}
where $A^{m,n}_k$ are constant coefficients. Taking into account
the formula \eeb \ we can represent $V^n$ also as polynomials of
the first current $V^{-1}=J$ and its derivatives. However, these
polynomials do not form a complete system.

For more clarification
we turn to the \ $U(1)$ \ group parameter \ $\varphi (z)$ \ in
terms of which the transformations laws \eeaa
\ take the form:
\eqn\eed{\delta ^n\varphi ={\cal U}^{-1}\delta
^n{\cal U}=\sum _{p,r\ge 1}(-)^{p-1}{\cal C}^{n+1,p-1}_{p,r}
\partial ^{p-r-1}k_n(x)\partial ^r{\cal P}^{n-p}(\partial \varphi),}
hence, the field $\varphi $ is always transformed
nonlinearly if $n>0$.

Rescaling the field $\varphi \rightarrow h^{-1}\varphi $ and
taking the
limit \ $h\rightarrow 0$ \ we derive from \eed \ the
area-preserving transformations law:
\eqn\eee{\widetilde \delta ^n\varphi =
\lim _{q\to 0}{q^{n+1}\over n+1}\delta ^n(q^{-1}\varphi )=
k_n(z)(\partial \varphi )^{n+1}}
which satisfy the \ $w_{\infty }$ \ Lie algebra:
\eqn\eeea{[\widetilde \delta ^m(k),\widetilde \delta ^n(h)]\varphi =
\widetilde \delta ^{m+n}(h_n\partial k_m-k_m\partial h_n)\varphi ,}
i.e. we have a contraction of the $W_{1+\infty }$--algebra to
the $w_\infty $--algebra.  We recall that the transformations
\eec \ and \eed
\ are only on-shell symmetry of the free scalar field action while the
$w_{\infty }$ transformations \ \eee \ are off-shell symmetry.

The Noether currents corresponding to the transformations \eee \ are
\eqn\eeeb{v^n(z)={1\over n+2}(\partial \varphi )^{n+2},}

If we turn to the group parameters as independent field
variables and assume the transformation law
\eqn\eeez{\delta ^n\varphi  =k_n(x)\partial ^n\varphi ,}
being  on-shell symmetry of \eeab \ we obtain a bosonic
realization of linear Gel'fand-Dikey algebra. The corresponding
conserved currents are given by
\eqn\eeeu{\tilde V^n =\partial \varphi \partial ^{n+1}\varphi ,}
These currents are generators of the same (up to redefinition) field
transformations \eeez . Hence, the higher spin symmetry
properties of the initial fermionic fields can be reproduced if
we consider the group parameters as independent field variables.

\subsec{Non-abelian Bosonization}

\noindent Similarly to the foregoing
cases we choose the higher spin conserved quantities
for the WZNW in the form:
\eqna\ewa
$$
\eqalignno{ V^n &
=tr\bigl(\partial g^{-1}\partial ^{n+1}g\bigr),&\ewa a \cr
J^n_a & =tr\bigl(\partial g^{-1}\partial ^{n}gt_a\bigr),&\ewa b
\cr }
$$
where $g$ takes its values on the group $G$, i.e.  $g\in G$. The
level parameter as well as the normalization constants are
omited in \ewa {} and what follows.  For definiteness we suppose
$G=SO(N)$, however our considerations are applicable for any
other group $G$.  In the present subsection we consider the
canonical conserved quantities only.  In paper \rZ \ it was
shown that these currents can be represented by the Fa\'a di
Bruno polynomials too, which show us that
\ewa{} is an appropriate bosonization of the higher spin fermionic
currents \ea{}. Indeed the currents \ewa{}  can be represented
in the form:
\eqn\ewb{\eqalign{ V^n &
=-tr\bigl(j(z)U^{n+1}(z)\bigr), \cr
J^n_a & =-tr\bigl(j(z)U_a^{n}(z)\bigr),
\cr }}
where the notations
\eqn\eaz{j(z)=U^1(z)=g^{-1}\partial g(z), \qquad \
U^n(z)=g^{-1}\partial ^{n}g(z), \qquad \
U^n_a(z)=g^{-1}\partial ^ngt_a=U^n(z)t_a}
are introduced. This coincide with the Sugawara
construction for $n=0$, hence for $n>0$, the formulas \ \ewb{}
can be considered as extended Sugawara constructions.

It is easy to verify that the following recursive relations hold:
\eqn\eeg{U^{n+1}=(j+\partial )U^n=DU^n, }
 where \ $D=J+\partial $ \ is a
matrix covariant derivative.  Starting from \ $U^0=I$, where \
$I$ \ is the identity matrix and \ $U^0_a=t_a$,  we obtain:
\eqn\eef{U^n=D^n.I=P^n(j),
\qquad \ U^n_a=D^nt_a=P^n(j)t_a,}
which are the matrix
generalizations of the Fa\'a di Bruno polynomials. Inserting \eef \ into
\eaz \ we get:
\eqn\eaaa{\eqalign{ V^n(z) &
=-tr\bigl(j(z)D^{n+1}(z)\bigr)=-tr\bigl(jP^{n+1}(j)\bigr), \cr
J^n_a(z) &=-tr\bigl(j(z)D^{n}t_a\bigr)
= -tr\bigl(jP^n(j)t_a\bigr).\cr }}

Rescaling the current \ $j \rightarrow
q^{-1}j$ \  in the limiting case $q\rightarrow 0$ we obtain:
\eqna\eaab
$$
\eqalignno{v^n & =\lim _{q\to 0}{q^{n+2}\over n+2}V^n(q^{-1}j)=
{1\over n+2}tr(j)^{n+2}, & \eaab a\cr
j^n_a & =\lim _{q\to 0}{q^{n+1}\over n+1}J_a^n(q^{-1}j)=
{1\over n+1}tr\bigl((j)^{n+1}t_a\bigr), & \eaab b\cr }
$$
The currents  \eaab{a} coincide with the area-preserving
conserved currents. In order to assure that the currents
\eaab{b}  have a Noether character we should introduce higher
 rank isotopic tensor currents (see \rZ \ and
\rRZ ).

In the case when $g$ is defined over the orthogonal group, i.e.\
$g\in SO(N)$, hence \ $g^{-1}=g^T$ \ which allows us to
write   \eqna\eka
$$
\eqalignno{(1-(-)^m)V^m(x)&=\sum _{p\ge 1}(-)^{m-p}{m\choose p}
\partial ^pV^{m-p},&\eka a\cr
(1-(-)^m)J_a^m(x)&=\sum _{p\ge 1}(-)^{m-p}{m\choose p}
\partial ^pJ_a^{m-p},&\eka b\cr }
$$
Notice, that the equality \eka{b} \ takes place only for
$(t_a)^T=-t_a$, while the second rank symmetric tensor currents $J^n_{ab}$
satisfy the Eq.\eka{a}. From \eka{}  we obtain
\eqna\ekb
$$
\eqalignno{V^{2m+1}&=\sum _{p=0}^mB^{2m+1}_{2p+1}\partial
^{2p+1}V^{2(m-p)},&\ekb a\cr
J_a^{2m+1}&=\sum _{p=0}^m\widetilde B^{2m+1}_{2p+1}\partial
^{2p+1}J_a^{2(m-p)},&\ekb b\cr }
$$
where $B^m_p$ are coefficients which can be determined from \eka{}.
The explicit form of \ekb{}  for the currents with lowest spin
 reads:
\eqn\ekc{\eqalign{V^{-1}&=trg^T\partial g\equiv 0, \cr
V^1&={1\over 2}\partial V^0,\cr
V^3&={3\over 2}\partial V^2-{1\over 4}\partial ^3V^0,\cr
V^5&={1\over 2}(-5\partial V^4+5\partial ^3V^2-\partial ^5V^0),
\cr
&\vdots \cr
J_a^1&=tr\partial g^Tt_a\partial g\equiv 0, \cr
J_a^3&=\partial J_a^2, \cr
J_a^5&=2\partial J_a^4-\partial ^3J_a^2. \cr
&\vdots \cr }}
Notice that the coefficients in these relations for the currents
\ $V$ \ and \ $J_a$ \ with coinciding index differ because the
current \ $V^n$ \ has a conformal spin \ $s=n+2$ \ while the
conformal spin of the current \ $J$ \ is \ $s=n+1$

The formulas \ekb{} (\ekc ) show us that the odd spin $U(1)$ currents
$V^{2m+1}$ are represented as derivatives from the
underlying even spin currents, while the even spin currents
$J^{2m+1}_a$ (the current $J^0_a$ has conformal spin equal to
one) are represented in terms of derivatives from the underlying
odd spin currents $J^{2l}_a$. Consequently, in the case of the
$SO(N)$ WZNW model the $U(1)$ currents with even spin are
independent, while only the isotopic vector currents with odd
spin are independent which corresponds to the $SO(N)$ free
fermionic model. By suitable redefinition of the currents \eaa{}
\ we can pass to a basis in which all dependent currents
vanish identically.

Notice, that in the current algebra for the $SO(N)$ currents
\ewa{b} \ as well as in the algebra of the corresponding field
transformation laws, which would be considered below, there
also arise the symmetric matrix generators \
$t_{ab}=\{t_a,t_b\}/2$ \ which as in the fermionic case complement
the $SO(N)$ algebra to the $GL(N)$ algebra.

We recall that the currents \ewa{} follow from the Noether
theorem as a consequence of the on-shell symmetry of the WZNW
action
with respect to the linear transformations (see Ref. \rZ ):
\eqna\eaac
$$
\eqalignno{\delta ^mg(z) & =k_m(z)\partial ^{m+1}g(z),&\eaac
a\cr
\widehat \delta ^mg(z) & =\alpha ^a_m(z)t_a\partial ^{m+1}g(z),
&\eaac b\cr }
$$
where $k_m , \alpha _m$ are arbitrary holomorphic functions.
These transformations satisfy the same Lie algebra \etr{} \  as the
corresponding fermionic transformations \eas{}. Taking
into account that $\delta ^mg^{-1}=-g^{-1}\delta ^mgg^{-1}$ it is
easy to verify that these transformations satisfy the same Lie
algebra as the transformations for $\delta ^mg$.  Notice, that
due to the dimensionlessness of the scalar field $g$ the
transformations \eaa{a} can be considered twofold: ones as
space-time conformal transformations and others as $U(1)$ gauge
transformations. From this property the Sugawara
construction for the stress-energy tensor follows. In the present paper
the transformations \eaa{a} are considered as $U(1)$ gauge
transformations.     However, as in the abelian case the
currents \ewa{}  do not form invariant space with respect to the
transformations \eaac{}  (see Refs.
\rZ , \rRZ ). Applying the
equal-time Poisson brackets we derive the fields transformations
generating from the currents\ewa{}:
\eqn\eab{\eqalign{\delta ^mg(x) &
=\int dyk_m(y)\{V^m(y),g(x)\}_{PB}=
{4\pi \over N}\biggl\{-k_m(x)\partial ^{m+1}g(x) \cr & +
g(x)\sum _{p\ge 1,q\ge 0}(-)^{p-1}{\cal C}^{m+1,p-1}_{p,q}
\partial ^qk_m\partial
^{p-q-1}\bigl(\partial g^{-1}\partial ^{m-p+1}g\bigr)\biggr\}, \cr
\widehat \delta ^mg(x) &
=\int dy\alpha^a_m(y)\{J_a^m(y),g(x)\}_{PB}
={4\pi \over N}\biggl\{-\alpha^a_m(x)t_a\partial ^{m}g(x) \cr & +
g(x)\sum _{p\ge 1,q\ge 0}{\cal C}^{m,p-1}_{p,q}
\partial ^q\alpha^a_m\partial
^{p-q-1}\bigl(\partial g^{-1}t_a\partial ^{m-p}g\bigr)\biggr\}, \cr }}
We use the Poisson bracket defined in \rW \ in which the
higher spin current variation is inserted by
\eqn\eac{\eqalign{\delta J^m_A & =tr\biggl(\partial \delta
g^{-1}\partial^mgA+
\partial g^{-1}\partial ^m\delta gA\biggr) \cr &
=tr\biggl(-\partial ^mgAg^{-1}\partial (\delta gg^{-1})+
\sum _{p\ge 1}{m \choose p}
\partial ^{m-p}gA\partial g^{-1}\partial ^p(\delta gg^{-1})\biggr)
\cr }}
and
\eqn\eacz{(\delta gg^{-1})(x)(\delta gg^{-1})(y)=
{4\pi \over N}I\otimes I\epsilon (x-y).}
Notice, that the anti-holomorphic coordinate \ $\bar z$ \ appears as
evolution (time) variable.

{\bf Observation:} The variations \eab \ for $m=0$ are linear
and coincide with the ordinary conformal and gauge (affine
Kac-Moody) transformations, while for \ $m>0$ \ nonlinear terms
arise. The appearance of these nonlinear terms is in contrast with
the initial transformations \eaac{} and the corresponding spinor
transformations \eaa{} .

We note, that the transformations \eab \ derived by means of the
Poisson bracket are on-shell symmetry of the WZNW action too,
which allow us to apply the Noether theorem again. The nonlinear
terms in \eab \ generate nonlinear currents of type $tr(jU^mU^n)
\ {\rm and} \ tr(jt_aU^mt_bU^n)$.  By repeatedly applying  the
Noether theorem we obtain set of nonlinear currents:
\eqn\esz{V^{n_1,n_2,\dots ,n_l}=tr\bigl(U^{n_1}U^{n_2}\dots
U^{n_l}\bigr), \qquad (l=1,2,\dots ).}
Taking into account that $U^0=I$ \ and \ $U^1=j=g^{-1}\partial
g$ from \esz \ we obtain $V^{n_1,\dots ,n_r,0,\dots
,0}=V^{n_1,\dots ,n_r}$ if $(r=2,3,\dots )$ and
$V^n=V^{1,n+1,0,\dots ,0}$ if $r=1$. It is straightforward to verify that
the extended currents space $\{V^{n_1,n_2,\dots ,n_l}\}$ is
invariant with respect to the transformations \eab . For
instance, the transformation law for the currents \esz \ with
respect to the linear transformations \eaac{a} \ reads
\eqn\esza{\eqalign{\delta ^mV^{n_1,n_2,\dots ,n_l}
& =\sum _{q=0}^{l-1}\Bigl\{-k_mV^{n_1,\dots ,n_q,m,n_{q+1},\dots
,n_l} \cr
& +\sum _{p\ge 0}{n_q\choose p}\partial
^pk_mV^{n_1,\dots ,n_{q-1}n_q+m-p+1,n_{q+1},\dots
,n_l}\Bigr\}.\cr }}

To have an invariant current space with respect to the extended
affine transformations too, we include  symmetric isotopic
tensor currents $J^{n_1,n_2,\dots ,n_l}_{a_1,\dots ,a_p},
(l,p=1,2,\dots ; \ p<l)$. For example the explicit form of the
rank $2$ tensor currents is:
\eqn\eszb{J^{n_1,\dots ,n_l}_{a,b}=\sum _{q,p=0,q\ne
p}^ltr\Bigl(U^{n_1}\dots U^{n_p}t_a\dots U^{n_q}t_b\dots
U^{n_l}\Bigr).}
It is straightforward to check that the set of
all tensor currents form an invariant space.

The same result we find if we consider the algebra of
nonlinear infinitesimal transformations \eab .

\newsec{Bosonic Current Algebra}

\subsec{Abelian Model}

\noindent First we consider the abelian model whose currents are
given by \eea . Notice, that for our purposes it is more
convenient to consider the $U(1)$ principal chiral model for
which  the group element ${\cal U}$ are independent field
variables.  In this case we derive the following current algebra
of the improved currents \eea :
\eqn\era{\eqalign{\{V^m(x),V^n(y)\}_{PB} &
= \delta V^m(x)\delta V^n(y) \cr
& =4\pi \sum _{p,q,r\ge 1,s\ge 0}{\cal C}^{m+2,p-1}_{p,r}
{\cal C}^{n+2,q}_{q,s}\partial ^{q-s}V^{m-p}\partial ^{p-r-1}V^{n-q}
\partial ^{r+s}\delta (x-y), \cr }}
where we used
\eqn\erb{\delta V^m(x)=\sum _{p\ge 1}{m+2\choose p}
V^{m-p}\partial ^p\bigl({\cal U}^{-1}\delta {\cal U}\bigr)}
and
\eqn\erc{{\cal U}^{-1}(x)\delta {\cal U}(x)
{\cal U}^{-1}(y)\delta {\cal U}(y)=4\pi \epsilon (x-y).}

The r.h.s.\ of \era \ contains linear as well nonlinear terms
which can be seen by rewriting \era \ in the following
equivalent form:
\eqn\eraa{\eqalign{\{V^m(x),V^n(y)\}_{PB} &
= \delta V^m(x)\delta V^n(y) \cr
& =4\pi \biggl\{\sum _{p,q,r=1,s= 0}^{m+1,n+1,p-1,q}
{\cal C}^{m+2,p-1}_{p,r}{\cal C}^{n+2,q}_{q,s} \cr
& \times \partial ^{q-s}V^{m-p}\partial ^{p-r-1}V^{n-q}
\partial ^{r+s}\delta (x-y)\cr
& +
\sum _{q,r\ge 1}^{n+1,m+1}{\cal C}^{n+2,m+1}_{q,r}
\partial ^{m-r+1}V^{n-q}\partial ^{q+r}\delta (x-y) \cr
& +\sum _{p,s\ge 1}^{m+1,n+1}{\cal C}^{m+2,n+2}_{p,s}
\partial ^{n-s+2}V^{m-p}\partial ^{p+s-1}\delta (x-y) \cr
& +\partial ^{m+n+3}\delta (x-y)\biggr\}.\cr }}

 The first term in the r.h.s.\ of \eraa
\ gives the nonlinear Gel'fand-Dikey algebra while the second and
third terms represent the linear part of the same algebra.
Although we are dealing with a classical theory there
 appear central terms in \eraa \ for any \ $m,n$.  We
note, that the algebra \eraa \ becomes pure linear only in the case \
$m=n=0$ \ i.e. for the Virasoro subalgebra.  Indeed, inserting \
$m=0$ \ into \eraa \ we obtain
\eqn\erba{\eqalign{\{V^0(x),V^n(y)\}_{PB}
& =4\pi
\sum _{q\ge 1}{n+2\choose q}\biggl(2\sum _{s=0}^q{m+2\choose r}
\partial ^{q-s}V^{-1}V^{n-q}\partial ^{s}\delta (x-y) \cr
& +\partial \bigl(V^{n-q}\partial ^q\delta (x-y)\bigr)\biggr), \cr }}
which if \ $n>0$ \ also contain nonlinear terms. Nonlinear terms
disappear only if we set \ $n=0$:
\eqn\erbc{\{V^0(x),V^0(y)\}_{PB}=\biggl(2V^0\partial +\partial V^0
+2\partial V^{-1}\partial +3V^{-1}\partial ^2+\partial ^3\biggr)
\delta (x-y),}
i.e. \ $V^0$ \ is a quasi-primary field. Consequently, after the
bosonization the linear current algebra transforms into
nonlinear ones with central terms on the classical level. The
latter is obvious for the improved quantities.

Here we also give the corresponding algebra for the canonical
currents \eua \ which reads:
\eqn\erd{\eqalign{\{\widetilde V^m(x),\widetilde V^n(y)\}_{PB} & =
\partial \widetilde V^{m-1}\widetilde V^{n-1}\delta (x-y)+
\widetilde V^{m-1}\widetilde V^{n-1}\partial \delta (x-y) \cr
& -\sum _{q\ge 1,s\ge 0}{\cal C}^{n+1,q}_{q,s}
\partial ^{q-s}\widetilde V^{m-1}\partial \widetilde V^{n-q}
\partial ^s\delta (x-y) \cr
& +\sum _{p\ge 1,r\ge 0}{\cal C}^{m+1, p}_{p,r}
\widetilde V^{m-p}\partial ^{p-r}\widetilde V^{n-1}
\partial ^r\delta (x-y) \cr &
+\sum _{p,q\ge 1,r,s\ge 0}{\cal C}^{m+1,p-1}_{p,r}{\cal C}^{n+1,q}_{q,s}
\partial ^{q-s}\widetilde V^{m-p}\partial ^{p-r}
\widetilde V^{n}
\partial ^{r+s}\delta (x-y), \cr }}
without a central term contrary to the improved current algebra
\era{} .

It is easy to check that the currents \eeeb \ satisfy the
same \ $w_{\infty }$ \ algebra \eeea \ with respect to the Poisson
brackets as the corresponding
infinitesimal transformations \eee.

\subsec{Non-abelian case}

\noindent In the non-abelian case we derive the following current
algebra: \eqna\ere
$$
\eqalignno{\{V^m(x),V^n(y)\}_{PB} & =
tr\biggl(\partial U^{m+1}U^{n+1}\biggr)\delta (x-y)+
tr\biggl(U^{m+1}U^{n+1}\biggr)\partial \delta (x-y) \cr
& -\sum _{q\ge 1,s\ge 0}{\cal C}^{n+1,q}_{q,s}
tr\biggl(\partial ^{q-s}U^{m+1}\partial W^{n-q+1}\biggr)
\partial ^s\delta (x-y) \cr
& +\sum _{p\ge 1,r\ge 0}{\cal C}^{m+1,p}_{p,r}
tr\biggl(W^{m-p+1}\partial ^{p-r}U^{n+1}\biggr)
\partial ^r\delta (x-y) \cr
& +\sum _{p,q \ge 1 r,s\ge 0}{\cal C}^{m+1,p-1}_{p,q}
{\cal C}^{n+1,q}_{q,s}
tr\biggl(\partial ^{q-s}W^{m-p+1}\partial ^{p-r-1}W^{n+1}\biggr)\cr
&\times \partial ^{r+s}\delta (x-y),&\ere a\cr
\{V^m(x),J^n_a(y)\}_{PB} & =
tr\biggl(\partial U^{m+1}Y^{n}_a\biggr)\delta (x-y)+
tr\biggl(U^{m+1}Y^{n}_a\biggr)\partial \delta (x-y) \cr &
-\sum _{q\ge 1,s\ge 0}{\cal C}^{n,q}_{q,s}
tr\biggl(\partial ^{q-s}U^{m+1}\partial j^{n-q}_a\biggr)
\partial ^s\delta (x-y) \cr
& +\sum _{p\ge 1,r\ge 0}{\cal C}^{m+1,p}_{p,r}
tr\biggl(W^{m-p+1}\partial ^{p-r}Y^{n}_a\biggr)
\partial ^r\delta (x-y) \cr
& +\sum _{p,q\ge 1 r,s\ge 0}{\cal C}^{m+1,p-1}_{p,r}
{\cal C}^{n,q}_{q,s}
tr\biggl(\partial ^{q-s}W^{m-p+1}\partial ^{p-r-1}j^{n}_a\biggr)\cr
&\times\partial ^{r+s}\delta (x-y),&\ere b\cr
\{J^m_a(x),J^n_b(y)\}_{PB} & =
tr\biggl(\partial Y^{m}_aY^{n}_b\biggr)\delta (x-y)+
tr\biggl(Y^{m}_aY^{n}_b\biggr)\partial \delta (x-y) \cr
& -\sum _{q\ge 1,s\ge 0}{\cal C}^{n,q}_{q,s}
tr\biggl(\partial ^{q-s}Y^{m}_a\partial j^{n-q}_b\biggr)
\partial ^s\delta (x-y) \cr
& +\sum _{p\ge 1,r\ge 0}{\cal C}^{m+1,p}_{p,r}
tr\biggl(j^{m-p}_a\partial ^{p-r}Y^{n}_b\biggr)
\partial ^r\delta (x-y) \cr
& +\sum _{p,q\ge 1,r,s\ge 0}{\cal C}^{m+1,p-1}_{p,r}
{\cal C}^{n+1,q}_{q,s}
tr\biggl(\partial ^{q-s}j^{m-p}_a\partial ^{p-r-1}j^{n-q}_b\biggr)\cr
&\times \partial ^{r+s}\delta (x-y),&\ere c\cr}
$$
where the following notations are used
\eqn\erg{\eqalign{W^n(x) & =\partial g^{-1}\partial ^{n}g(x)
=-U^1U^{n}(x), \cr
Y^n_a(x) & =g^{-1}\partial ^ng(x)t_a, \cr
j^n_a(x) & =\partial g^{-1}\partial ^ng(x)t_a
=-U^1Y_a^n(x). \cr }}
We remark that
\eqn\erh{\eqalign{V^n(x) & =trW^{n+1}, \cr
J^n_a(x) & = trj^n_a(x). \cr }}

We note that in the l.h.s.\ of the current algebra \ere{}
nonlinear terms (with respect to the matrixvalued currents \erg
) appear, which  in the general case are impossible to be
represented (lineally or nonlinearly) in terms of the currents
\erh \ only.  Consequently, to have a closed algebra additional
nonlinear currents of type  \esz \ and\eszb \ should be included
(in both caseslinear and nonlinear algebra).

Taking into account
\eqn\erl{\partial U^n(x)=U^{n+1}-U^1U^n}
we derive the following generalized Laibnitz formula:
\eqn\erm{\eqalign{\partial ^lU^m(x)
& =\sum _{s\ge 0}{l\choose s}
\sum _{r_1\ge 0}{s\choose r_1} \cr
& \times \sum _{r_2\ge 0}{s-r_1\choose r_2}\dots
\sum _{r_{k-1}\ge 0}{s-r_1-\dots -r_{k-1}\choose r_k}\cr
 & \times U^{r_k}U^{r_{k-1}}\dots U^{r_2}U^{r_1}U^{m+l-s}
\delta _{r_1+r_2+\dots +r_k-s,0}. \cr }}
Hence, the r.h.s. of \ere{} \ can be
represented in terms of the currents \esz \ and \eszb . As
was pointed out above the
currents \esz \ and \eszb \ are also Noetherian and are
generated by the nonlinear part of the transformations \eab .
The nonlinear currents \esz \ satisfy the following current
algebra:
\eqn\ern{\eqalign{\{V^{\{m\}_k}(x),V^{\{n\}_l}\}_{PB}
& =\sum _{CP(m)}\sum _{CP(n)}
tr\Bigl(U^{m_1}(x)U^{m_2}(x)\dots U^{m_{k-1}}(x) \cr
& \times U^{n_1}(x)U^{n_2}(x)\dots U^{n_{l-1}}(x) \cr
& \times \sum _{p,q\ge 1,r,s\ge 0}{\cal C}^{m,p-1}_{p,r}
{\cal C}^{n,q}_{q,s}
\partial ^{q-s}U^{m-p}\partial ^{p-r-1}U^{n-q}\Bigr)
\partial ^{r+s-1}\delta (x-y), \cr }}
where $\{m\}_k=(m_1,\dots ,m_k)$ and $CP(m)$ denotes the
cyclic permutations of $m_1,\dots ,m_k$.

Taking into account \erm \ the r.h.s.\  of \ern \ can be
rewritten only in
terms of the currents \esz \ (without derivative terms). This
allow us to conclude that currents \esz \ satisfy a closed
linear algebra if \ $l$ \ runs to $\infty $. In the case of
isotopic currents \eszb \ one obtains closed linear algebra if
currents with increasing isotopic indices up to
\eqn\erp{J^{m_1,\dots ,m_k}_{a_1,\dots,a_k}
=tr\Bigl(U^{m_1}t_{a_1}\dots U^{m_2}t_{a_2}\dots
U^{m_k}t_{a_k}\Bigr), \qquad (k=1,2,\dots )}
are included. Consequently, the currents \esz , \eszb , \erp \
satisfy a closed linear affine algebra if $t_a\in
{\cal SL(N)}$ also in the case $g\in SO(N)$.

Notice, that we obtained an algebra which is a triple
extension of the semi-direct product of the Virasoro algebra and
the affine Kac-Moody algebra. This extension is carried out with
respect to: the conformal spin, the degree of nonlinearity and the
isotopic spin. Let us recall that there exists
possibility for the algebra \ere{} to be
considered as a nonlinear algebra for matrix valued
currents. We point out, that there are several possibilities to
obtain particular extended algebras from the general
construction considered above.  A simple example of the
contraction procedure is presented below.

Inserting \eef \ into \esz ,
\eszb , \erp \ and after redefinition  $j\rightarrow
q^{-1}j$ in the limiting case $q\rightarrow 0$ we get:
\eqna\elli
$$
\eqalignno{v^n(x)
&=\lim_{q\to 0}{q^{n+2}\over n+2}
V^{n_1,n_2,\dots ,n_k}={1\over n+2}trj^{n+2},&\elli a\cr
v_a^n(x)&=\lim_{q\to 0}{q^{n+1}\over n+1}
J_a^{n_1,n_2,\dots ,n_k}={1\over n+1}tr\bigl(t_aj^{n+1}\bigr),
&\elli b\cr
v_{a_1,a_2,\dots ,a_l}^n(x)&=\lim_{q\to 0}{q^{n+1}\over n+1}
J_{a_1,a_2,\dots ,a_l}^{n_1,n_2,\dots ,n_k},&\elli c\cr}
$$
where $n=\sum _{r=0}^kn_r$ \ and \ $l\leq k$. We note, that if
\ $1<l<k$ \ there are various possibilities for re-arrangering the
matrix generators $t_a$ and the currents $U^n$ in the formula
\elli{a} .

With respect to the Poisson bracket the currents \elli{} \
satisfy the following algebra:
\eqna\eli
$$
\eqalignno{\{v^m(x),v^n(y)\}_{PB}&=(m+1)\partial v^{m+n}(x)
\delta (x-y)+(m+n+2)v^{m+n}\partial \delta (x-y),&\eli a\cr
 \{v^m(x),v_b^n(y)\}_{PB}&={1\over n+1}\sum _{p=0}^n
tr\biggl(\partial \bigl(j^{m+1}\bigr)j^pt_bj^{n-p}\biggr)\delta
(x-y) \cr
&+(m+n+1)v^{m+n}_b\partial \delta (x-y),&\eli b\cr
 \{v^m_a(x),v_b^n(y)\}_{PB}&={1\over (m+1)(n+1)}\sum _{p=0}^m
\biggl\{\sum _{q=0}^n
tr\biggl(\partial \bigl(j^pt_aj^{m-p}\bigr)j^pt_bj^{n-p}\biggr)\delta
(x-y) \cr
&+tr\biggl(j^{n+p-q}t_aj^{m+q-p}t_b\biggr)\partial \delta (x-y)\biggr\}.
&\eli c \cr }
 $$
We note, that the terms of type \ $tr\bigl(\partial
(j^m)Aj^n\bigr)$ \ can be represented as a total derivative
only if  $A=I$, otherwise, the term
$tr(j^{m-k-1}\partial j j^{k}t_aj^n)\equiv
tr(j^{m-k-1}U^2j^kt_aj^n)-tr(j^{m+1}t_aj^n)$ also is included.
Consequently, it can be concluded that only the currents
\elli{a} satisfy a closed linear algebra \eli{a}
coinciding with the area-preserving \ $w_{\infty
}$ \ algebra. The r.h.s. of \eli{b,c} \  shows that the
isotopic vector and generally isotopic tensor currents do not
form closed algebra if the currents containing the
derivatives of \ $j$ \ are not included, i.e.\ the currents of
type \erp .

\newsec{Conclusions}

\noindent The main results obtained in the present article is
that when the bosonization procedure is applied some higher
spin symmetry properties are changed. For the free
fermionic model it is well known that the higher spin currents
and their generating transformations form the same Lie algebra
--the $DOP(S^1)$ algebra (coinciding with the linear part of the
Gel'fand-Dickey algebra) which agree with the Noether theorem.
This result take place for the both abelian and non-abelian
models.  In the corresponding bosonic models (the free single
scalar field model and the WZNW model), we start from the same
type ($DOP(S^1)$) linear fields transformations and its
corresponding bilinear higher spin currents. By cyclic
application of the Noether theorem nonlinear terms arise in the
field transformations laws, hence in the corresponding currents.
The scalar field current algebra contains both the linear and
non-linear Gel'fand-Dickey algebras.  Moreover, in the general
non-abelian case  current algebra is closed only if the terms
with an increasing degree of nonlinearity are included. These
nonlinear terms involve currents with increasing isotopic spin
too. As a result we obtain triple extended algebra.  This
current algebra can be considered as well as a nonlinear
extended algebra of type $\widehat W_\infty $ if we consider
matrix valued currents. All this allows us to conclude that some
difference appears between the higher spin symmetry properties
of the non-abelian free fermionic model and the corresponding
WZNW model. We recall that on the ordinary symmetry level both
models are completely equivalent.

\bigskip\centerline{\bf Acknoledgements}

\noindent The Author would like  to thank Prof. Abdus Salam, the
International Atomic Energy Agency and the United Nations
Educational, Scientifical  and Cultural Organization for the
hospitality in the International Centre for  Theoretical Physics
in Trieste, where the final part of this paper was done. It is a
pleasure to thank also Profs. L. Bonora, I. T. Todorov and D. Tz.
Stojanov and Drs. G. M. Sotkov and R. Paunov for valuable
discussions in the various stage of this work.



\appendix {A}{}

\noindent Here as an example is considered the
group $SL(2,R)$. For this group the Gauss decomposition reads:
\eqn\epa{g(x)=\Psi \Phi \Delta ,}
where
\eqn\epb{\Psi =\pmatrix {1&0\cr \psi &1\cr } ,\qquad
\Phi =\pmatrix {e^{-\phi }&0\cr 0&e^{\phi } \cr },\qquad
\Delta =\pmatrix{1&\chi \cr 0&1\cr }.}
Here $\psi, \  \phi \ {\rm and} \  \chi $ \ are real scalar
dimensionless fields.

Inserting \epa \ and \epb \  into \eaz \ we obtain
\eqn\epc{U^m=U^m_+\sigma _-+U^m_-\sigma _++
U^m_3\sigma _3+U^m_0\sigma _0,}
where \eqna\epd
$$
\eqalignno{U^m_+&=\partial ^m\psi +2\psi A_m-e^{2\phi }\psi ^2
\partial ^m\chi \cr
&+\sum _{p\ge 1}{m\choose p}\biggl(\partial
^{m-p}\psi P^p(\partial \phi )-
e^{2\phi }\psi \partial ^{m-p}\psi \partial ^p\chi -
e^{2\phi }\psi ^2\partial ^p\chi P^{m-p}(\partial \phi ) \cr
&-\sum _{q\ge 1}{m-p\choose q}e^{2\phi }
\psi \partial ^{m-p-q}\psi \partial ^p\chi
P^q(\partial \phi)\biggr),&\epd a\cr
U^m_-&=e^{2\phi }\partial ^m\chi +
\sum _{p\ge 1}{m\choose p}e^{2\phi }\partial ^p\psi
P^{m-p}(\partial \phi ),&\epd b\cr
U^m_3&=-A_m+e^{2\phi }\psi \partial ^m\chi \cr
&+{1\over 2}\sum _{p\ge 1}{m\choose p}\biggl(
e^{2\phi }\partial ^{m-p}\psi \partial ^p\chi +
2e^{2\phi }\psi \partial ^p\chi P^{m-p}(\partial \phi ) \cr
&+\sum _{q\ge 1}{m-p \choose q}e^{2\phi }
\partial ^{m-p-q}\psi \partial ^p\chi
P^q(\partial \phi)\biggr),&\epd c\cr
U^m_0&=B_m+{1\over 2}\sum _{p\ge 1}{m\choose p}\biggl(
e^{2\phi }\partial ^{m-p}\psi \partial ^p\chi \cr
&+\sum _{q\ge 1}\pmatrix{m-p \cr q \cr }e^{2\phi }
\partial ^{m-p-q}\psi \partial ^p\chi
P^q(\partial \phi)\biggr).&\epd d\cr }
$$
Here  $A_m$  and \ $B_m$ are defined by
\eqn\epda{\Phi ^{-1}\partial ^m\Phi =-A_m\sigma _3+B_m\sigma _0}
$\sigma _0=I$ and $\sigma _{\pm }, \sigma _0$ are matrix
generators of the $SL(2,R)$ algebra:
\eqn\epdc{\sigma _3=\pmatrix{1&0\cr 0&-1\cr }, \qquad
\sigma _+=\pmatrix{0&1\cr 0&0\cr },\qquad
\sigma _-=\pmatrix {0&0\cr 1&0\cr }.}
{}From \epda \ it holds that
\eqn\epdd{\eqalign{A_{m+1}&=\partial A_m+\partial \phi B_m, \cr
B_{m+1}&=\partial B_m+\partial \phi A_m, \cr }}
where $A_0=0, B_0=1$. From \epdd \ it follows that the sum $A_m+B_m$
satisfies the condition \eab   , i.e. we have
\eqn\epde{P^m(\partial \phi )=A_m(\phi )+B_m(\phi ),}
where $P^m$ are the Fa\' a di Bruno polynomials.
Taking into account that $U^1=j(x)$ from \epc \ we obtain
\eqn\epcr{\eqalign{j(x)&=g^{-1}\partial g(x) \cr
& =(\partial\psi +2\psi \partial \phi -e^{2\phi }\psi ^2\partial
\chi )\sigma _-
+e^{2\phi }\partial \chi \sigma _++(-\partial \phi +
e^{2\phi }\psi \partial \chi )\sigma _3. \cr }}
Now, it is easy to compute
\eqn\epea{j^2=\bigl((\partial \phi )^2+
e^{2\phi }\partial \chi \partial \psi \bigr)I=2TI,}
where $I$ is the identity matrix and $T$ is the stress-energy
tensor. Then, from
\epc\  and \epd{}  we have:
\eqn\epf{\eqalign{j^{2k} & =2T^kI,\cr
j^{2k+1} & =2T^kj(x) \qquad {\rm for} k=0,1,\dots . \cr }}
Inserting \epf \ into \elli{}  we find:
\eqn\eph{\eqalign{& v^{2k}=T^k,\cr
& v^{2k+1}=0.\cr }}

Consequently, all the odd spin $w_\infty $ conserved quantities
\elli{}  vanish in the case of $SL(2,R)$ WZWN model, while the
even spin quatities are a product of stress-energy tensor.

{}From \epc \ we get
\eqn\eepha{V^m={1\over 2}trU^{m+2}=U_0^{m+2}}
which for $m>-1$ do not vanish identically. Consequently, the
formulas \epc , \epd{}  and \eepha \ show us that the $W_\infty $
currents with any spins do not vanished identically.

Next we show that the currents $V^{m,n,\dots }$ can not be
represented in terms of the currents $V^m$, i.e.
\eqn\ephb{V^{m,n,\dots }\ne V^mV^n\dots }
Indeed inserting \epc \ in the product $U^mU^n$ we get
\eqn\ephc{\eqalign{U^mU^n&=(U^m_+U^n_3-U^m_3U^n_++
U^m_+U^n_0+U^m_0U^n_+)\sigma _- \cr
&+(U^m_3U^n_--U^m_-U^n_3+
U^m_-U^n_0+U^m_0U^n_-)\sigma _+ \cr
&\biggl({1\over 2}(U^m_-U^n_+-U^m_+U^n_-)+
U^m_3U^n_0+U^m_0U^n_3\biggr)\sigma_3 \cr
&\biggl({1\over 2}(U^m_-U^n_++U^m_+U^n_-)+
U^m_3U^n_3+U^m_0U^n_0\biggr)\sigma_0. \cr }}
Consequently, in general $tr(U^mU^n)\ne tr(U^m)tr(U^n)$ if
$m,n\ge 0$.  This allows us to conclude that we have a closed
current algebra only if we consider an extended currents space
in which higher spin nonlinear currents are included.
This is also the case of WZNW model for which the gauge fixing
$U^m_\pm =0, U_3^m=-A^m, U^m_0=B^m$ is impose. Substituting in
\ephc \ we obtain:
\eqn\ezz{tr(U^mU^n)=2(A^mA^n+B^mB^n)\ne cB^mB^n}
because according to \epdd \ we have $A^m\ne B^m$.

\listrefs
\bye